\documentclass[10pt,a4paper]{article}
\pdfoutput=1

\usepackage[T1]{fontenc}
\usepackage{mathptmx}            
\usepackage{microtype}
\usepackage[margin=2.5cm]{geometry}
\usepackage{amsmath,amssymb,amsthm}
\usepackage{mathtools}
\usepackage{bm}
\usepackage{authblk}
\usepackage[numbers]{natbib}
\usepackage[colorlinks=true,linkcolor=black,citecolor=black,urlcolor=blue]{hyperref}

\hypersetup{
  pdftitle={Polynomial integrals and Poisson structures of the Pais-Uhlenbeck oscillator},
  pdfauthor={Ilmar Gahramanov and Gulmammad Mammadov}
}

\title{\bfseries Polynomial integrals and Poisson structures\\ of the Pais-Uhlenbeck oscillator}

\author[1,2]{Ilmar Gahramanov\thanks{E-mail: ilmar.gahramanov@bogazici.edu.tr}}
\author[3]{Gulmammad Mammadov\thanks{E-mail: gmammadov@ada.edu.az}}
\affil[1]{Department of Physics, Bo\u{g}azi\c{c}i University, Istanbul, T\"urkiye}
\affil[2]{Center for Mathematics and its Applications, Khazar University, Baku, Azerbaijan}
\affil[3]{School of Information Technologies and Engineering, ADA University, Baku, Azerbaijan}

\date{}

\begin{document}
\maketitle

\begin{abstract}
\noindent
We classify the autonomous polynomial integrals of the nondegenerate
fourth-order Pais-Uhlenbeck oscillator. For a rational frequency ratio
$m:n$ in lowest terms, the two sector energies together with the real
and imaginary parts of a single ladder product generate all polynomial
integrals, subject to one defining relation. The smallest total degree
of an integral not expressible through the sector energies is $m+n$, and
this bound is attained. For an irrational ratio only polynomials in the
sector energies are conserved. In normal-mode variables the commensurate
case is the negative oscillator resonance, and the classification
realizes its invariant theory in the original higher-derivative
variables. Two compatible Poisson structures generate the same dynamics
with an indefinite and a positive-definite Hamiltonian. They induce
different brackets on the same physical generators. We show that
diagonal rescalings of the generators identifying the two abstract
Poisson algebras exist only when $m$ or $n$ is even, and that none of
them preserves the physical domain of nonnegative sector energies.
Finally, we express the generators in terms of the original coordinate
and its first three derivatives. The classical multiplier identity gives
their variational characteristics in closed form and shows that the
additional resonant integrals cannot arise from point variational
symmetries.

\end{abstract}

\section{INTRODUCTION}

The Hamiltonian treatment of higher-derivative theories goes back to
Ostrogradsky \cite{Ostrogradsky}, and the resulting Hamiltonian is
generically unbounded from below (see e.g.\ \cite{Woodard} for a
review). A basic example is
the fourth-order oscillator introduced by Pais and Uhlenbeck in 1950
\cite{PU}. Its equation of motion factorizes into two harmonic-oscillator
operators, and for distinct real frequencies the free classical motion
is bounded, although the Ostrogradsky Hamiltonian is indefinite. Its
quantization and alternative Hamiltonian descriptions have been studied
extensively, see e.g.\ \cite{MannheimDavidson,BolonekKosinski} and
references therein.

In terms of the two decoupled modes, additional integrals of motion
appear naturally: for commensurate frequencies a combination of the two
phases is constant. Bolonek and Kosi\'nski sketched the construction of
such an integral \cite{BolonekKosinski}, and Andrzejewski, Galajinsky,
Gonera and Masterov showed that the model is maximally superintegrable
whenever the frequency ratios are rational \cite{AndrzejewskiGGM}. This
is more general than the odd-frequency sequence required for the
$l$-conformal Newton-Hooke symmetry studied in the same
work.\footnote{For the fourth-order oscillator this sequence corresponds
to the ratio $3:1$.}

Our purpose is to give a direct account of the polynomial integrals of
the fourth-order model and to connect them with its higher-derivative
Lagrangian. Complex ladder functions provide a short proof of the
complete polynomial classification, including the minimal degree $m+n$
and the irrational case. \footnote{Ladder and factorization techniques for superintegrable systems are reviewed in \cite{BHKN}. The rational anisotropic oscillator goes back to Jauch and Hill \cite{JauchHill}. See also \cite{MPW}.} We compare two compatible Poisson structures that
generate the same vector field with Hamiltonians $H_1-H_2$ and $H_1+H_2$.
Their brackets differ on the fixed physical generators. Diagonal
rescalings of the generators identify the two abstract Poisson algebras
only when $m$ or $n$ is even, and none of them preserves the physical
domain $H_1,H_2\ge0$. We then express the generators in the original
coordinate and give their generalized variational symmetries in closed
form. This complements \cite{CobanGK}, where the variational symmetries
of second-order Lagrangians are generated by point transformations, and
gives a direct proof that the additional resonant generators cannot be
Noether charges of point transformations.

These results should be seen against the well-established theory of
oscillator resonances. The positive-definite $m:n$ resonance is
discussed in \cite{CushmanBates}. The invariant rings for both $m:n$ and
$m:-n$ resonances, together with the inequalities describing the image
of the invariant map, are given in \cite[Sec.~0.4.1]{EfstathiouThesis}
(see also \cite{EfstathiouBook}), and the associated Poisson geometry is
studied in \cite{HolmVizman}. The normal-mode form of the
Pais-Uhlenbeck Hamiltonian identifies its commensurate case with the
negative resonance. Our classification therefore realizes this invariant
theory in the original higher-derivative variables, where it acquires a
Lagrangian and variational interpretation.

Further Hamiltonian formulations and
symmetries of the Pais-Uhlenbeck oscillator are discussed in
\cite{Andrzejewski2014,MasterovOdd,MasterovSUSY,LieGhostFree}. Elbistan
showed, by means of the Eisenhart-Duval lift, that the model describes a
particle in a circularly polarized periodic gravitational wave and
derived its conserved charges from the symmetries of the wave, which
include Carroll symmetry \cite{Elbistan2022}. Elbistan and Andrzejewski found further realizations
of the model, used its bi-Hamiltonian character, and obtained additional
integrals from the conformal Newton-Hooke symmetry
\cite{ElbistanAndrzejewski}. Bi- and tri-Hamiltonian formulations were
used to construct ghost-free representations \cite{FelskiFringTurnerTri},
and an interacting version was shown to be integrable \cite{FelskiFring}.
On the quantum side, coherent states \cite{PramanikGhosh}, ghost-free
quantization \cite{FringTairaTurner} and the equal-frequency case
\cite{FringMarquetteTaira} were studied. Here we restrict ourselves to
distinct frequencies and autonomous classical integrals.

The paper is organized as follows. In section~\ref{sec:modes}, we recall
the normal-mode form of the Ostrogradsky Hamiltonian. In
section~\ref{sec:factorization}, we construct the additional integrals,
and in section~\ref{sec:classification} we classify all polynomial
integrals. Section~\ref{sec:algebra} contains the Poisson algebra and the
comparison of the two Hamiltonian structures and their physical domains. In
section~\ref{sec:original}, we write the integrals in terms of $q$ and
identify the corresponding variational symmetries. We conclude with a
discussion of possible extensions.

\section{HAMILTONIAN DESCRIPTION AND NORMAL MODES}
\label{sec:modes}

We start with the Lagrangian
\begin{equation}
L=\frac12\left[\ddot q^2-(\omega_1^2+\omega_2^2)\dot q^2
+\omega_1^2\omega_2^2q^2\right],
\label{lag}
\end{equation}
which leads to the equation of motion
\begin{equation}
\left(\frac{d^2}{dt^2}+\omega_1^2\right)
\left(\frac{d^2}{dt^2}+\omega_2^2\right)q=0.
\label{eqPU}
\end{equation}
Throughout the paper we assume $\omega_1>\omega_2>0$ and set
$\Delta=\omega_1^2-\omega_2^2$.\footnote{The equal-frequency case requires
a separate treatment, since the normal-mode transformation below becomes
singular. Complex conjugate frequency pairs, considered in
\cite{BolonekKosinski}, are likewise outside the present scope.} The
Ostrogradsky variables are
\begin{equation}
Q_1=q,\qquad Q_2=\dot q,\qquad P_2=\ddot q,\qquad
P_1=-(\omega_1^2+\omega_2^2)\dot q-\dddot q,
\end{equation}
with the Hamiltonian
\begin{equation}
H=P_1Q_2+\frac12P_2^2+\frac12(\omega_1^2+\omega_2^2)Q_2^2
-\frac12\omega_1^2\omega_2^2Q_1^2
\label{OstH}
\end{equation}
and the Poisson bracket
\begin{equation}
\{f,g\}=\sum_{j=1}^2\left(
\frac{\partial f}{\partial Q_j}\frac{\partial g}{\partial P_j}
-\frac{\partial f}{\partial P_j}\frac{\partial g}{\partial Q_j}\right),
\qquad \dot F=\{F,H\}.
\label{PB}
\end{equation}

The linear canonical transformation
\begin{align}
x_1&=\frac{P_2+\omega_2^2Q_1}{\sqrt\Delta},&
p_1&=-\frac{P_1+\omega_1^2Q_2}{\sqrt\Delta},\nonumber\\
x_2&=\frac{P_2+\omega_1^2Q_1}{\sqrt\Delta},&
p_2&=\frac{P_1+\omega_2^2Q_2}{\sqrt\Delta}
\label{diagmap}
\end{align}
brings the Hamiltonian to the form
\begin{equation}
H=H_1-H_2,\qquad H_j=\frac12(p_j^2+\omega_j^2x_j^2),
\label{diagH}
\end{equation}
where the sector energies $H_1, H_2\ge0$ are separately conserved. The
joint level sets $H_1=E_1>0$, $H_2=E_2>0$ are two-tori, which is why the
free motion is bounded even though the level sets of the indefinite
Hamiltonian are noncompact. In terms of the normal modes
\begin{equation}
u=\ddot q+\omega_2^2q,\qquad
v=\ddot q+\omega_1^2q,\qquad
\ddot u=-\omega_1^2u,\quad \ddot v=-\omega_2^2v,
\label{uv}
\end{equation}
one has
\begin{equation}
x_1=\frac{u}{\sqrt\Delta},\quad p_1=\frac{\dot u}{\sqrt\Delta},\qquad
x_2=\frac{v}{\sqrt\Delta},\quad p_2=-\frac{\dot v}{\sqrt\Delta},
\qquad
H_1=\frac{\dot u^2+\omega_1^2u^2}{2\Delta},\quad
H_2=\frac{\dot v^2+\omega_2^2v^2}{2\Delta}.
\label{uvmap}
\end{equation}

\section{LADDER FUNCTIONS AND ADDITIONAL INTEGRALS}
\label{sec:factorization}

Each sector energy factorizes as
\begin{equation}
H_1=B^+B^-,\qquad B^\pm=\frac{\omega_1x_1\mp ip_1}{\sqrt2},
\qquad
H_2=A^+A^-,\qquad A^\pm=\frac{\omega_2x_2\mp ip_2}{\sqrt2},
\label{ladders}
\end{equation}
where $B^\pm$ and $A^\pm$ are complex conjugate pairs. One finds
\begin{align}
\{B^+,B^-\}&=i\omega_1,& \{A^+,A^-\}&=i\omega_2,\nonumber\\
\{H_1,B^\pm\}&=\mp i\omega_1B^\pm,&
\{H_2,A^\pm\}&=\mp i\omega_2A^\pm,
\label{sectorbrackets}
\end{align}
while brackets between different sectors vanish. Since $H=H_1-H_2$,
\begin{equation}
\{H,B^\pm\}=\mp i\omega_1B^\pm,\qquad
\{H,A^\pm\}=\pm i\omega_2A^\pm,
\label{HBA}
\end{equation}
i.e.\ functions with the same label evolve with opposite phases.

Let the frequency ratio be rational,
\begin{equation}
\frac{\omega_1}{\omega_2}=\frac mn,\qquad m>n>0,\qquad \gcd(m,n)=1.
\label{ratio}
\end{equation}
Then the products
\begin{equation}
X^\pm=(B^\pm)^n(A^\pm)^m
\label{integral}
\end{equation}
satisfy $\{H,X^\pm\}=\mp i(n\omega_1-m\omega_2)X^\pm=0$. The resonance
condition attaches the smaller exponent $n$ to the higher frequency
$\omega_1$. This recovers the construction of \cite{AndrzejewskiGGM}. For the positive-definite
Hamiltonian $H_1+H_2$ the conserved products pair opposite labels
instead. The real functions
\begin{equation}
X=\frac{X^++X^-}{2},\qquad
Y=\frac{X^+-X^-}{2i}
\label{realXY}
\end{equation}
are homogeneous polynomials of degree $m+n$ and satisfy
\begin{equation}
X^2+Y^2=X^+X^-=H_1^nH_2^m.
\label{relation}
\end{equation}
Writing $B^\pm=\sqrt{H_1}e^{\pm i\theta_1}$ and
$A^\pm=\sqrt{H_2}e^{\pm i\theta_2}$ on the region $H_1, H_2>0$, equation
\eqref{HBA} gives $\dot\theta_1=\omega_1$, $\dot\theta_2=-\omega_2$, and
\begin{equation}
X=R\cos\psi,\qquad Y=R\sin\psi,\qquad
R=H_1^{n/2}H_2^{m/2},\qquad \psi=n\theta_1+m\theta_2.
\label{phase}
\end{equation}
Thus $X$ and $Y$ carry, in addition to the amplitudes, one relative-phase
combination.

\section{CLASSIFICATION OF POLYNOMIAL INTEGRALS}
\label{sec:classification}

Let us show that the construction above exhausts the polynomial
integrals, i.e.\ polynomials in the phase-space coordinates without
explicit time dependence.\footnote{Oscillators also admit explicitly
time-dependent conserved quantities linear in the coordinates and
momenta. For higher-derivative Lagrangians such integrals arise naturally
from variational symmetries, see e.g.\ \cite{CobanGK}. The degree used
below is the total degree in positions and momenta. Since
\eqref{diagmap} is linear, it is the same in the Ostrogradsky variables
and in $(q,\dot q,\ddot q,\dddot q)$.} Since the ladder functions are
invertible linear combinations of the canonical variables, any polynomial
is a combination of monomials
\begin{equation}
M=(B^+)^a(B^-)^b(A^+)^c(A^-)^d,
\qquad a,b,c,d\in\mathbb Z_{\ge0},
\end{equation}
on which the Hamiltonian acts diagonally,
\begin{equation}
\{H,M\}=-i\big[(a-b)\omega_1-(c-d)\omega_2\big]M.
\label{weight}
\end{equation}
Since the monomials are linearly independent, every term of a conserved
polynomial must satisfy $(a-b)\omega_1=(c-d)\omega_2$. For an irrational
ratio this forces $a=b$, $c=d$, so only polynomials in $H_1,H_2$ are
conserved. For a rational ratio the condition reads $(a-b)m=(c-d)n$, and
coprimality gives $a-b=kn$, $c-d=km$ with $k\in\mathbb Z$. Then
\begin{equation}
M=H_1^bH_2^d(X^+)^k\quad (k\ge0),\qquad
M=H_1^aH_2^c(X^-)^{|k|}\quad (k<0).
\end{equation}
Hence every conserved polynomial is a polynomial in $H_1,H_2,X^\pm$.
Each monomial with $k\ne0$ has total degree at least $|k|(m+n)$.
Distinct ladder monomials cannot cancel, so every polynomial integral
outside $\mathbb C[H_1,H_2]$ has degree at least $m+n$. The real
generators $X$ and $Y$ attain this bound.

To see that~\eqref{relation} is the only relation, note that with
$z_+z_-=h_1^nh_2^m$ any polynomial in formal generators $h_1,h_2,z_\pm$
reduces to $f_0(h_1,h_2)+\sum_{k\ge1}\big(f_k(h_1,h_2)z_+^k
+g_k(h_1,h_2)z_-^k\big)$. After substitution, different terms give
distinct ladder monomials, so such an expression vanishes only if all
coefficients vanish. Complex conjugation fixes $H_1,H_2$ and
interchanges $X^+$ and $X^-$, so the real polynomial integrals are
generated by $H_1,H_2,X,Y$, and we obtain
\begin{equation}
\mathcal I_{\mathbb R}\simeq
\frac{\mathbb R[h_1,h_2,x,y]}
{(x^2+y^2-h_1^nh_2^m)},
\label{ring}
\end{equation}
where $h_1,h_2,x,y$ represent $H_1,H_2,X,Y$. For $H_1+H_2$ the same
argument applies with $A^+$ and $A^-$ interchanged and gives the ring of
the $m:n$ resonance \cite{CushmanBates}, isomorphic to~\eqref{ring} as a
commutative ring. Finally,
\eqref{phase} gives $dH_1\wedge dH_2\wedge dX=-R\sin\psi\,dH_1\wedge
dH_2\wedge d\psi$ and $dH_1\wedge dH_2\wedge dY=R\cos\psi\,dH_1\wedge
dH_2\wedge d\psi$, so where both modes are present at least one of $X,Y$
is independent of $H_1,H_2$. Thus the invariant map has rank three on
the open dense set $H_1, H_2>0$, and the system is maximally
superintegrable. On the boundary the rank can drop. For instance, at
$H_2=0$ the differentials $dH_2$, $dX$ and $dY$ all vanish, since $m\ge2$,
and the same happens at $H_1=0$ when $n\ge2$.

\section{POISSON ALGEBRA AND THE HAMILTONIAN SIGN}
\label{sec:algebra}

\subsection{The algebra of integrals}

From~\eqref{sectorbrackets} and the Leibniz rule one finds
$\{H_1,X^\pm\}=\mp in\omega_1X^\pm$, $\{H_2,X^\pm\}=\mp im\omega_2X^\pm$
and
\begin{equation}
\{X^+,X^-\}=i\left(n^2\omega_1H_1^{n-1}H_2^m
+m^2\omega_2H_1^nH_2^{m-1}\right).
\label{complexalgebra}
\end{equation}
With $\Omega=n\omega_1=m\omega_2$, the real generators obey
\begin{align}
\{H_1,H_2\}&=0,\nonumber\\
\{H_1,X\}=\{H_2,X\}&=\Omega Y,\nonumber\\
\{H_1,Y\}=\{H_2,Y\}&=-\Omega X,\nonumber\\
\{X,Y\}&=-\frac{\Omega}{2}H_1^{n-1}H_2^{m-1}
(nH_2+mH_1).
\label{realalgebra}
\end{align}
In particular, $H_1-H_2$ commutes with every generator, and
$X^2+Y^2-H_1^nH_2^m$ is a Casimir of the formal algebra which vanishes in
this realization.

\subsection{Two compatible Poisson structures}

Besides the canonical bracket, consider the modified bracket
\begin{equation}
\{f,g\}_{\mathrm{alt}}=
\frac{\partial f}{\partial x_1}\frac{\partial g}{\partial p_1}
-\frac{\partial f}{\partial p_1}\frac{\partial g}{\partial x_1}
-\frac{\partial f}{\partial x_2}\frac{\partial g}{\partial p_2}
+\frac{\partial f}{\partial p_2}\frac{\partial g}{\partial x_2},
\label{altbracket}
\end{equation}
which differs from the canonical one by the sign of the second sector.
Both Poisson tensors are constant, so every linear combination of them
is again a Poisson tensor. The brackets are therefore compatible. The
associated recursion operator is the constant involution
$R=\mathrm{diag}(1,1,-1,-1)$ in the coordinates $(x_1,p_1,x_2,p_2)$, so
the pencil generates no hierarchy beyond the two Hamiltonians
$H_1\mp H_2$. Moreover,
\begin{equation}
\{F,H_1-H_2\}=\{F,H_1+H_2\}_{\mathrm{alt}}
\label{bihamiltonian}
\end{equation}
for every $F$. In other words, the vector field of the Pais-Uhlenbeck
oscillator is Hamiltonian with respect to two Poisson structures, with
the indefinite Hamiltonian $H_1-H_2$ and with the positive-definite
Hamiltonian $H_1+H_2$, respectively. This is a concrete instance of the
bi-Hamiltonian character of the model
\cite{LieGhostFree,ElbistanAndrzejewski,FelskiFringTurnerTri}. Hamiltonian
structures of the model admitting quadratic Hamiltonians were classified
earlier in \cite{BolonekKosinski}.

The two descriptions generate the same vector field and hence have the
same integrals, generated by $H_1,H_2,X,Y$ with the
relation~\eqref{relation}. Their brackets on these fixed generators
differ. In particular,
\begin{align}
\{H_1,H_2\}_{\mathrm{alt}}&=0,\nonumber\\
\{H_1,X\}_{\mathrm{alt}}&=\Omega Y,&
\{H_1,Y\}_{\mathrm{alt}}&=-\Omega X,\nonumber\\
\{H_2,X\}_{\mathrm{alt}}&=-\Omega Y,&
\{H_2,Y\}_{\mathrm{alt}}&=\Omega X.
\label{altmixed}
\end{align}
Since
$\{A^+,A^-\}_{\mathrm{alt}}=-i\omega_2$, one finds
\begin{align}
\{X^+,X^-\}_{\mathrm{alt}}&=i\left(n^2\omega_1H_1^{n-1}H_2^m
-m^2\omega_2H_1^nH_2^{m-1}\right),\nonumber\\
\{X,Y\}_{\mathrm{alt}}&=-\frac{\Omega}{2}H_1^{n-1}H_2^{m-1}(nH_2-mH_1).
\label{positivealgebra}
\end{align}
The comparison is summarized in table~\ref{tab:comparison}. The map
$p_2\mapsto-p_2$ turns the modified bracket into the canonical one and
interchanges $A^+$ and $A^-$, so the right column is the familiar
algebra of the positive-definite oscillator with integrals
$(B^\pm)^n(A^\mp)^m$ \cite{CushmanBates}. The canonical bracket with
$H_1-H_2$ gives the negative-resonance algebra, whose symplectic leaves
are unbounded, in contrast to the bounded leaves of the positive
resonance \cite{HolmVizman}. Thus the two descriptions agree on the
equations of motion while assigning different brackets to the same
physical observables.

\begin{table}[t]
\centering
\small
\setlength{\tabcolsep}{4pt}
\renewcommand{\arraystretch}{1.5}
\begin{tabular}{lcc}
\hline
 & canonical bracket, $H_1-H_2$ & modified bracket, $H_1+H_2$\\
\hline
vector field & \multicolumn{2}{c}{same}\\
polynomial integrals & \multicolumn{2}{c}{$H_1,H_2,X,Y$,\quad
$X^2+Y^2=H_1^nH_2^m$}\\
$\{H_2,X\}$ & $\Omega Y$ & $-\Omega Y$\\
$\{H_2,Y\}$ & $-\Omega X$ & $\Omega X$\\
$\{X,Y\}$ & $-\frac{\Omega}{2}H_1^{n-1}H_2^{m-1}(nH_2+mH_1)$
& $-\frac{\Omega}{2}H_1^{n-1}H_2^{m-1}(nH_2-mH_1)$\\
level sets of the Hamiltonian & noncompact & compact\\
\hline
\end{tabular}
\caption{Two compatible Hamiltonian structures of the Pais-Uhlenbeck
oscillator. Each column uses its indicated bracket. The same physical
generators have different brackets. The last row refers to nonempty
Hamiltonian level sets in the original real phase space.}
\label{tab:comparison}
\end{table}

\subsection{Abstract algebras and the physical invariant domain}
\label{sec:physicaldomain}

The distinction between the brackets on fixed observables should be
separated from the question of abstract Poisson-algebra isomorphism.
For example, if $m$ is even, the change of generators
\begin{equation}
(\widetilde H_1,\widetilde H_2,\widetilde X,\widetilde Y)
=(H_1,-H_2,X,Y)
\label{evenmmap}
\end{equation}
preserves the defining relation~\eqref{relation} and converts the
canonical brackets~\eqref{realalgebra} into
\eqref{altmixed} and \eqref{positivealgebra} for the tilded generators.
If $n$ is even, the same conclusion follows from
\begin{equation}
(\widetilde H_1,\widetilde H_2,\widetilde X,\widetilde Y)
=(-H_1,H_2,X,-Y).
\label{evennmap}
\end{equation}
Both statements follow by direct substitution. In either case the
indefinite central element $H_1-H_2$ is, up to an overall sign, the sum
of the tilded sector generators.

These two maps are in fact the only diagonal rescalings with this
property. Requiring
$(\widetilde H_1,\widetilde H_2,\widetilde X,\widetilde Y)
=(\alpha H_1,\beta H_2,\gamma X,\delta Y)$ to carry~\eqref{realalgebra}
into~\eqref{altmixed} and~\eqref{positivealgebra}, the mixed brackets
force $(\alpha,\beta,\delta)=(1,-1,\gamma)$ or $(-1,1,-\gamma)$, and the
bracket $\{X,Y\}$ together with~\eqref{relation} then requires
$\gamma^2=\alpha^n\beta^m>0$, that is $m$ even in the first case and $n$
even in the second. Since $\gcd(m,n)=1$, both may be odd, and the ratio
$3:1$ is of this type. For $m$ and $n$ both odd no such rescaling exists,
and we leave open whether the two algebras are isomorphic by a nonlinear
change of generators. These examples show why different bracket formulas
alone do not establish abstract non-isomorphism.

The physical realization carries additional information. The image of
the invariant map is the semialgebraic set
\begin{equation}
\mathcal S=\left\{(h_1,h_2,x,y)\in\mathbb R^4:
h_1\ge0,\ h_2\ge0,\ x^2+y^2=h_1^nh_2^m\right\}.
\label{physicaldomain}
\end{equation}
The sign changes in~\eqref{evenmmap} and~\eqref{evennmap} do not preserve
this domain. More generally, every diagonal rescaling found above has
$\alpha\beta=-1$ and therefore reverses the sign of one sector
generator. They therefore do not identify the two physical
realizations with nonnegative sector energies. This role of the
inequalities is also explicit in the reduction of positive and negative
resonances \cite[Sec.~0.4.1]{EfstathiouThesis}. The distinction between
compact levels of $H_1+H_2$ and noncompact levels of $H_1-H_2$ in
table~\ref{tab:comparison} concerns these physical realizations.

\section{INTEGRALS IN TERMS OF \texorpdfstring{$q$}{q} AND VARIATIONAL SYMMETRIES}
\label{sec:original}

Using~\eqref{uvmap}, the complex generators can be written directly in
terms of $q$ and its first three derivatives,
\begin{equation}
X^\pm=(2\Delta)^{-(m+n)/2}
(\omega_1u\mp i\dot u)^n(\omega_2v\pm i\dot v)^m.
\label{Xq}
\end{equation}
On the solutions $u=a_1\cos(\omega_1t+\varphi_1)$,
$v=a_2\cos(\omega_2t+\varphi_2)$ this gives
$X^\pm\propto a_1^na_2^m e^{\pm i(n\varphi_1-m\varphi_2)}$, so the
conserved combination of physical phases is the familiar
$n\varphi_1-m\varphi_2$. Indeed, for $a_1,a_2>0$ one has
$\theta_1=\omega_1t+\varphi_1$ and $\theta_2=-(\omega_2t+\varphi_2)$ on
these solutions, so that $\psi=n\varphi_1-m\varphi_2$. The plus sign
in~\eqref{phase} thus reflects the opposite orientation of the second
canonical angle.

We now identify the variational symmetries of~\eqref{lag} associated
with these integrals. Following \cite{CobanGK}, we consider
$q\to q+\epsilon\,Q$, but
allow the characteristic $Q$ to depend on derivatives of $q$. It is a
variational symmetry if $\delta L=dG/dt$ identically for some $G$.
Integrating by parts,
\begin{equation}
\delta L=E\,Q+\frac{d}{dt}\left(QP_1+\dot QP_2\right),\qquad
E=q^{(4)}+(\omega_1^2+\omega_2^2)\ddot q+\omega_1^2\omega_2^2q,
\label{deltaL}
\end{equation}
where $P_1,P_2$ are the Ostrogradsky momenta, so the Noether charge is
$C=QP_1+\dot QP_2-G$, which is the case of no time reparametrization in
the general expression of \cite{CobanGK}. Conversely, let
$I(q,\dot q,\ddot q,\dddot q)$ be an autonomous integral. Since
\eqref{eqPU} is solved for $q^{(4)}$, every first integral admits a
representative of this differential order, so this is no restriction
\cite{AncoBluman}. Its time
derivative contains $q^{(4)}$ only through $(\partial I/\partial\dddot
q)\,q^{(4)}$, and since $\dot I$ vanishes for arbitrary
$q,\dot q,\ddot q,\dddot q$ on solutions, one has the identity
$dI/dt=(\partial I/\partial\dddot q)\,E$. In the language of
conservation laws, $\partial I/\partial\dddot q$ is the integrating
factor, or multiplier, of $I$, and the identity is the standard statement
that the multiplier of a first integral is its derivative with respect to
the highest derivative \cite{AncoBluman,Olver}. Choosing
\begin{equation}
Q_I=-\frac{\partial I}{\partial\dddot q}=\{q,I\},
\label{charQ}
\end{equation}
equation~\eqref{deltaL} gives $\delta L=\frac{d}{dt}(Q_IP_1+\dot
Q_IP_2-I)$, so $I$ is the Noether charge of $Q_I$. For this fixed
off-shell representative of the charge, the characteristic is unique
among those depending on derivatives of order at most three: if $I$ is
also the charge of $Q'$, then $(Q'-Q_I)E=0$ identically, and the
coefficient of $q^{(4)}$ gives $Q'=Q_I$. Charges may be shifted by
constants without changing their characteristics. The uniqueness
statement fixes the charge as a function of
$(q,\dot q,\ddot q,\dddot q)$.

The construction is tied to the canonical structure. The identity
$Q_I=-\partial I/\partial\dddot q$ follows from~\eqref{deltaL} alone and
does not refer to any bracket, but the second equality in~\eqref{charQ}
holds for the canonical bracket only. Since $q=(x_2-x_1)/\sqrt\Delta$ by
\eqref{diagmap}, the modified bracket~\eqref{altbracket} gives
$\{q,I\}_{\mathrm{alt}}\ne\{q,I\}$ in general. The variational
formulation based on the Lagrangian~\eqref{lag} therefore singles out the
canonical one of the two Hamiltonian structures of
section~\ref{sec:algebra}, even though both generate the same flow.

For the sector energies, $Q_{H_1}=-\dot u/\Delta$ and
$Q_{H_2}=-\dot v/\Delta$ depend on $\dddot q$. Among their linear
combinations, precisely the multiples of
$Q_{H_1-H_2}=\dot q$ are point characteristics, corresponding to time
translation. For the additional integrals,
\begin{equation}
Q^\pm=\mp i\,(2\Delta)^{-\frac{m+n}{2}}
(\omega_1u\mp i\dot u)^{n-1}(\omega_2v\pm i\dot v)^{m-1}
\left[m(\omega_1u\mp i\dot u)-n(\omega_2v\pm i\dot v)\right],
\label{Qpm}
\end{equation}
with $Q_X=(Q^++Q^-)/2$ and $Q_Y=(Q^+-Q^-)/(2i)$.

A point transformation has characteristic
$\eta(t,q)-\dot q\,\xi(t,q)$ and hence cannot depend on $\ddot q$ or
$\dddot q$. To prove that both $Q_X$ and $Q_Y$ have unavoidable
higher-derivative dependence, set $N=m+n\ge3$ and $r=\dddot q$.
On the subspace $\dot q=0$, one has $\dot u=\dot v=r$ and
$n\omega_1u-m\omega_2v=\Omega(u-v)=-\Omega\Delta q$. Expanding
\eqref{Qpm} in powers of $r$ therefore gives
\begin{equation}
\left.Q^+\right|_{\dot q=0}
=-(2\Delta)^{-N/2}i^{m-n}
\left[Nr^{N-1}-i\Omega\Delta(N-1)q\,r^{N-2}
+\mathcal R(q,\ddot q,r)\right],
\qquad \deg_r\mathcal R\le N-3.
\label{Qleading}
\end{equation}
The factor $i^{m-n}$ is either real or purely imaginary. Thus one of
$Q_X,Q_Y$ has a nonzero $r^{N-1}$ term and the other a nonzero
$q\,r^{N-2}$ term. Since $N\ge3$, both characteristics depend
nontrivially on $\dddot q$. By the uniqueness just established, neither
$X$ nor $Y$ can be the Noether charge of a point variational symmetry
of~\eqref{lag}.

As an example, consider $\omega_1=2\omega$, $\omega_2=\omega$, so that
$m=2$, $n=1$ and $\Delta=3\omega^2$. Up to normalization,
$X+iY\propto(2\omega u-i\dot u)(\omega v+i\dot v)^2=I+iJ$ with
\begin{align}
I&=2\omega\left[u(\omega^2v^2-\dot v^2)+\dot u\,v\,\dot v\right],&
Q_I&=-2\omega\left(\dot u\,v+v\,\dot v-2u\,\dot v\right),\nonumber\\
J&=4\omega^2u\,v\,\dot v-\dot u(\omega^2v^2-\dot v^2),&
Q_J&=-\left(4\omega^2uv-\omega^2v^2+2\dot u\,\dot v+\dot v^2\right),
\label{example}
\end{align}
where $u=\ddot q+\omega^2q$ and $v=\ddot q+4\omega^2q$. For the normalized
generators $X=(2\Delta)^{-3/2}I$, $Y=(2\Delta)^{-3/2}J$, the
algebra~\eqref{realalgebra} reduces to $\{X,Y\}=-\omega H_2(H_2+2H_1)$.
In this example $m$ is even, so the map~\eqref{evenmmap} applies. For the
ratio $3:1$, where $m$ and $n$ are both odd, no diagonal rescaling
identifies the two algebras, and the generators are quartic.

\section{CONCLUDING REMARKS}

We have given a direct classification of the autonomous polynomial
integrals of the nondegenerate fourth-order Pais-Uhlenbeck oscillator.
For a rational frequency ratio $m:n$ in lowest terms, the invariant
ring has four real generators and one defining relation, and the
minimal degree beyond the sector energies is $m+n$. For an irrational
ratio, the ring is generated by $H_1,H_2$ alone. These results realize
the established invariant theory of the negative oscillator resonance in
the original higher-derivative system, where it acquires a variational
interpretation.

The two compatible Hamiltonian structures have the same flow and
invariant ring but different brackets on the fixed physical generators.
The combinations $nH_2+mH_1$ and $nH_2-mH_1$ in $\{X,Y\}$ distinguish
these brackets. Diagonal rescalings identify the two abstract algebras
only when $m$ or $n$ is even, and none of them preserves the physical
domain $H_1,H_2\ge0$. For $m$ and $n$ both odd, which includes the ratio
$3:1$, it remains open whether the two abstract algebras are isomorphic
by a nonlinear change of generators. The distinction between the
invariant ring, its Poisson structure and its physical realization is
therefore essential.

In the Lagrangian description, the classical multiplier identity
$Q_I=-\partial I/\partial\dddot q$ gives an explicit variational
characteristic for each autonomous first integral, and its Hamiltonian
form $Q_I=\{q,I\}$ singles out the canonical bracket. For a fixed off-shell
charge, this characteristic is unique within the class of differential
order at most three. The dependence of $Q_X,Q_Y$ on $\dddot q$ proves
that the additional resonant generators require generalized variational
symmetries. Several extensions of this construction remain of interest.

For the ratio $3:1$, the Carroll charges of the corresponding
gravitational wave become combinations of the Newton-Hooke generators
\cite{Elbistan2022}, and the conformal symmetry supplies additional,
explicitly time-dependent integrals
\cite{AndrzejewskiGGM,ElbistanAndrzejewski}. It would be interesting to
express the quartic generators of this case in terms of these charges.
It would also be interesting to understand whether the generalized
symmetries of section~\ref{sec:original} correspond to higher-rank
Killing tensors of the gravitational-wave metrics obtained by the
Eisenhart-Duval lift \cite{Elbistan2022,ElbistanAndrzejewski}, and to
extend the analysis of \cite{CobanGK} systematically to generalized
symmetries of higher-derivative Lagrangians.

For more than two frequencies, the extension of the commensurate
construction is indicated in \cite{AndrzejewskiGGM}. The monomial argument
extends to the integer resonance lattice, although the generating sets
and their relations become more involved. Applying it to particular
higher-order Pais-Uhlenbeck models would also give explicit
higher-derivative realizations of their polynomial symmetry algebras.
Finally, a quantum extension must specify the Hamiltonian structure,
operator domains and ordering. The classical brackets determine the
leading semiclassical commutators, but the comparison of quantum
representations and spectra requires further analysis. Classically
equivalent formulations need not yield equivalent quantum theories, as
illustrated by the equal-frequency model in \cite{FringMarquetteTaira}.

\section*{ACKNOWLEDGMENTS}

Both authors belong to the scientific school of Emil Akhmedov and would
like to express their gratitude to him. The ideas and questions explored
in this paper have their roots in many discussions and lessons they had
with him. I. Gahramanov acknowledges the Istanbul Integrability and Stringy Topics Initiative (iStringy). 

\section*{FUNDING}

The work of I. Gahramanov was partially supported by the
Bo\u{g}azi\c{c}i University Research Fund under grant no.~Y26.
G. Mammadov received no specific funding for this work.

\section*{CONFLICT OF INTEREST}

The authors declare that they have no conflicts of interest.

\bibliographystyle{pannat}
\bibliography{PU_PAN}
\end{document}